# The effects of large climate fluctuations on forests as a Big Data approach

Costas A. Varotsos[1] and Vladimir F. Krapivin[2]

[1] Department of Environmental Physics and Meteorology, University of Athens, Athens, Greece;
[2] Kotelnikov Institute of Radioengineering and Electronics, Russian Academy of Sciences, Moscow, Russia.

**Abstract**

In the present study it is shown that significant impacts on forests can lead to large climate fluctuations at global level. This research takes into account various scenarios for the reconstruction of forest areas using the geo-ecological information-modeling system. This tool is described as a possible improvement of the Big Data approach. Its structure includes 24 blocks that combine a series of models and algorithms for the global processing of large data and their analysis. The key block of this tool is the global model of the climate-nature-society system.
.

## 1. Introduction

Present civilization development advances the problems of assessment and forecasting of the expected climate changes and related variations in habitat of humans and animals. In the first place, the beginning and expansion of dangerous natural processes leading to the loss of people and economic damages is one of the main problem of environmental monitoring data processing. The complexity of this problem is related to the heterogeneity and multi-formity of available information from various sources such as Earth monitoring systems and existing databases[1]. Indeed, human civilization must solve a more important problem of sustainable development between nature and society. What is needed is the investigation of reliable and efficient informational technology providing spatial scope of global and regional relationships between complex structures of relations within nature and society by considering possible constraints and multi-fold structures. According to the recent literature the fundamental problem lies in the conception of globalization and its understanding[2,3].

Interaction of humanity and nature is a function of a vast set of factors acting both in the human society and in the natural environment. The main problem of this interaction is the globalization of the anthropogenic impacts on the natural systems of population growth and the expansion of polluted areas. Existing Big Data approaches to the sustainable problem solution are mainly focused on traditional economic challenges[4-9] as well as on restricted environmental problems[10-16].

Nowadays, big data from satellites play a major role in climate change dynamics[17-20]. Over the last decades, atmospheric chemical composition such as carbon dioxide, ozone and other greenhouse gases has been systematically monitored by satellite and ground-based instruments[21-31]. Such measurements are now necessary to monitor extreme weather events by observing surface temperature and wind field, through advanced optical satellites and remotely-sensed ground-based instrumentation[32-40].

This paper provides a description of the approach and methodology to be used to solve the problem of sustainable development of the climate-nature-society system (CNSS) taking into account both natural and demographic processes.

Among the existing tools for environmental data visualization, the geophysical information system (GIS) is the most demanded approach to environmental monitoring data processing and representation. Basic GIS imperfection consists in that it does not focus on multi-pronged prognosis of monitoring objects. Important improvement of GIS technology was made more than ten years ago when geoecological information-modeling system (GIMS) was presented as a combination of GIS and modeling technology[41]. Key aspects of GIMS have been discussed in many publications[42-46].

## 2. Big Data approach and global sustainable development problems

With the development of society, the CNSS sustainable development problem becomes increasingly critical covering practically globe. Even the use of satellite environmental monitoring does not provide data that can help to assess the NSS characteristics with high reliability and, in particular, the prognosis of CNSS evolution. Big Data tools help solve limited economical problems but encounter difficulties when environmental considerations are taken into account when the nature protection problems are considered. Therefore, widening the Big Data tools with the provision of functions for the processing and analysis of super-large volumes of environmental information delivered from different sources irregularly at the time and fragmentary in space is a real problem. Solution of this problem is mainly realized with the use of global models oriented on the study of the CNSS restricted aspects, including climate and biospheric models that practically consume pre-historical structured data of restricted volume. The sustainability development problem of global CNSS for its solution requires the collection of data at unprecendented scale. Decisions that are based on existing global models of different environments, including biosphere, geosphere, atmosphere, etc., can not answer the main question: what is the optimal structure of global monitoring data that reflects different aspects of individual CNSS items and helps to overcome the un-removable uncertainties within the many areas.

As demonstrated recently the development of biogeochemical, biocenotic, hydrophysical, climatic and socio-economic processes taking place in the NSS inevitably requires a balanced criterion for information selection taking account the hierarchy of causative effects in the CNSS with the coordination of spatial digitization. Existing environmental monitoring systems such as the Earth Observing System (EOS) and the Global Ocean Observing System (GOOS) provide long-term global observations of the land surface, biosphere, atmosphere, solid Earth, and oceans and enable an improved understanding of the Earth as an integrated system. These systems have allowed for the synthesis of Earth Observing System Data and Information System (EOSDIS) that provides capabilities for managing data from various sources of different type, including satellites, aircraft, field measurements, and various other sources. The EOSDIS contains growing database in its ingestion of approximately 8.5 terabytes daily. The EOSDIS archive volume growths from last years from about 0.2 PB in 2000 to 14 PB in 2015. These data flows can be completed with the global and regional socioeconomic information from Socioeconomic Data and Applications Data Center (SEDAC) the catalog of which summarizes data available in 52 countries. The EOSDIS and SEDAC data combined processing can play a key role in the NSS sustainable development problem solution using the GIMS approach[45].

## 3. The GIMS as Big Data approach improvement

The generalized concept of GIMS is shown in Figures 1 and 2. The GIMS key item is a global NSS model[2,3]. Basic GIMS principles are:
- Integration, unification and coordination of big data fluxes delivered by the existing monitoring resources basing on the unique organizational and science-methodic basis;
- Coordination and compatibility of big data fluxes by using the unique coordinate-time system, common system for classification, coding, format and data structure; and
- Providing the independence of big data fluxes from ecosystem and state boundaries.

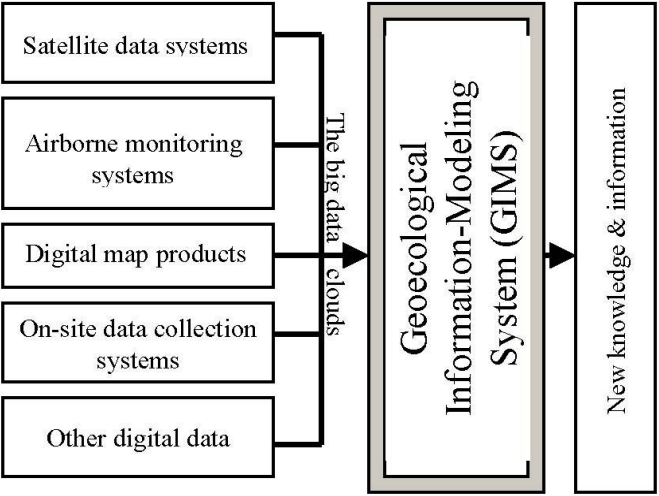

**Figure 1.** The GIMS as tool for integration of big data fluxes delivered by different monitoring systems and other data sources.

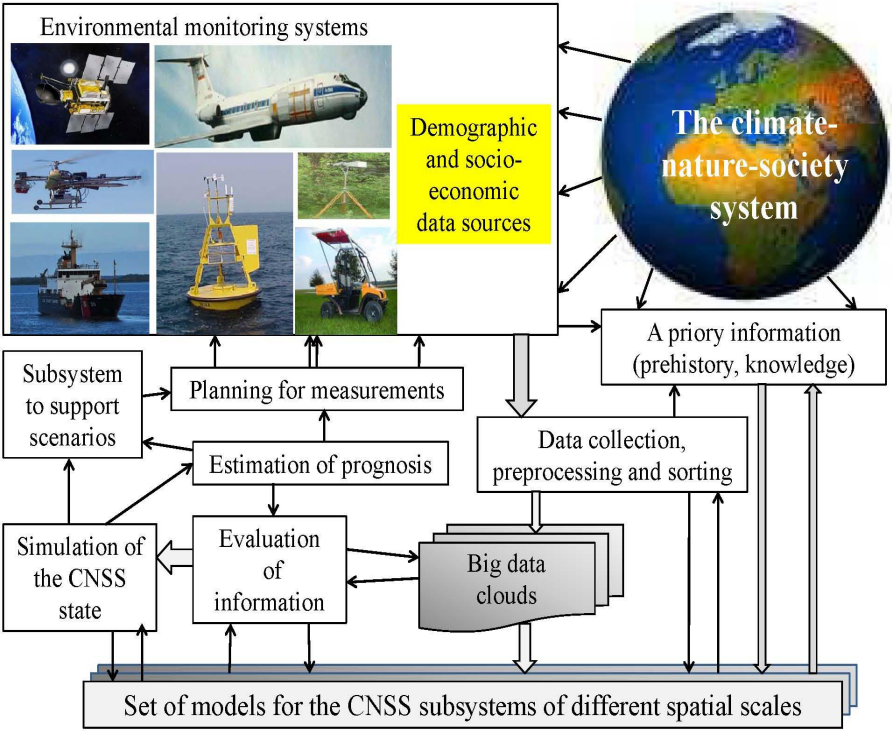

**Figure 2.** Conceptual block-diagram of the GIMS functional operations.

Construction of the GIMS is connected with consideration of the components of the biosphere, climate and social medium characterized by the given level of spatial hierarchy. Realization of the GIMS function is provided by its subsystems listed in Table 1. Basic block of the GIMS is the CNSS global simulation model (CNSSGSM), the structure of which is shown in Figure 3. The CNSSGSM assimilates big data fluxes considering complex of biospheric, climatic and socio-economic processes taking into account their space-temporal hierarchy. According to the procedure represented in Figure 2, the CNSSGSM structure is oriented on the adaptive functioning mode introducing the global model into a system of geoinformation monitoring[45]. The approach to the CNSSGSM synthesis is based on the two mathematical methods:

- Balance equations are used when knowledge of fluxes of matters and information between CNSS components are exhaustive.

- Evolutionary algorithm is applied when the build-up of an adequate balance model is, in principle, impossible because of the lack of information completeness and knowledge of environmental and socio-economic laws is insufficient.

**Table 1**. The GIMS functions and their short description.

| The GIMS function | The description of the function |
|---|---|
| Planning and analysis of big data clouds. | The analysis of the structure of the environmental data acquisition system using satellite data, flying laboratories, and mobile and stationary ground observation means as well as providing socio-economic information[45]. |
| Synchronous analysis of big data fluxes using space-time interpolation and extrapolation methods. | Retrieval of data and their reduction to the common time scale is performed. Global model parameters are determined. The thematic classification of big data is carried out and space-time combination is performed on measurements obtained from various types of sources[47]. |
| Evaluation of the state of the atmosphere. | The gas and aerosol composition of the near-earth atmospheric layer are provided and forecasting maps of their distribution are created[48,49]. |
| Evaluation of the state of the soil-plant covers. | Determining the structural topology of land cover revealing soil-plant formations in accordance with spatial resolution[45,47]. |
| Evaluation of the state of the water medium. | Simulation model is used for the hydrological processes taking into account seasonal changes of surface and river runoff, the influence of snow cover and permafrost and the regime of precipitation and evaporation[50,51]. |
| Modeling of the global biogeochemical cycles. | Mathematical models of global cycles of greenhouse gases are used taking into account the roles of soil-plant formations, World Ocean and geosphere as well as anthropogenic processes[52]. |
| Modeling of the photosynthesis. | In the oceans and vegetation layers the photosynthetic processes are described by proper mathematical models[53]. |
| Modeling of demographic processes. | Population dynamics is described by two models with consideration of the role of environmental and social factors. Models differ in mathematical approaches[47]. |
| Climate change modeling. | Climate change processes are described by simple functional models reflecting the roles of greenhouse gases and pollution of the atmosphere[54]. |
| Identification of causes of ecological and sanitary disturbances in the environment. | Investigation and identification of dangerous environmental processes is realized, including the detection and prognosis of tropical cyclones, floods and excessive atmosphere pollution[45]. |
| Intelligent support. | Software-mathematical algorithms are developed to provide the user with intelligent support in performing complex analysis of simulation experiment results[55]. |

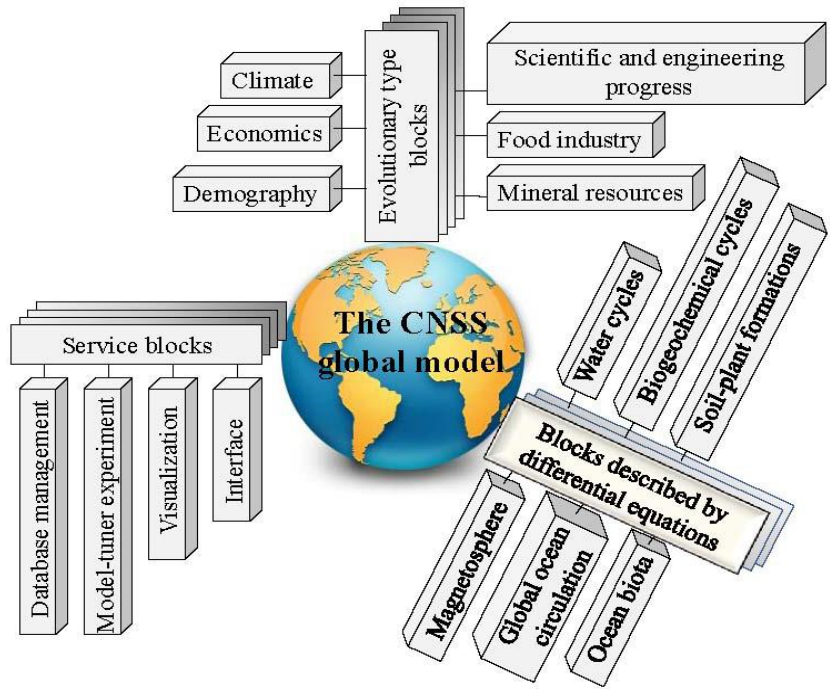

**Figure 3.** The structure of the CNSSGSM [46,47].

The GIMS-based method makes it possible to create a global monitoring system in which the CNSSGSM provides the entire system to be categorized as a class of subsystems with variable structures and makes the system adaptable to changes in natural and socio-economic processes. Figure 4 demonstrates the general structure of GIMS as the aggregate approaches, instruments and methods for the processing of structured and unstructured data characterized by big volumes and significant variety. GIMS allows the combined use of different approaches to the processing of big data fluxes that primarily solve a new decision-making process to optimize these fluxes at the expense of effective monitoring alternatives and data processing tools[50,51].

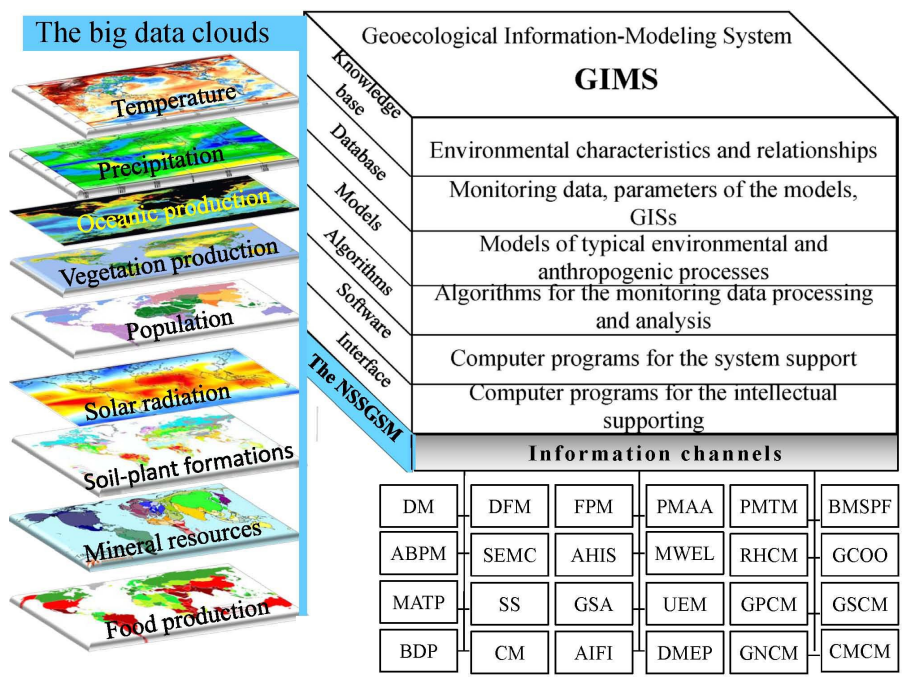

**Figure 4.** The GIMS/NSSGSM block-diagram. The explanation of abbreviations is given in Table 2.

**Table 2.** The GIMS/NSSGSM functional blocks.

| Block | Block functions |
|---|---|
| DM | Demographic model . |
| CM | Climate model . |
| CMCM | Coupled model of the carbon dioxide and methane cycles |
| GSCM | Global sulphur cycle model . |
| GCOO | Coupled model of global cycles of oxygen and ozone . |
| GNCM | Global nitrogen cycle model . |
| GPCM | Global phosphorus cycle model . |
| RHCM | Regional hydrological cycle model . |
| BMSPF | Biocenotic model of the soil-plant formations . |
| PMTM | Photosynthesis model for the tropical and moderate oceanic zones . |
| PMAA | Photosynthesis model for the Arctic and Antarctic zones of the World Ocean |
| ABPM | Arctic Basin pollution model . |
| MATP | Model of long-range atmospheric transport of the pollutants |
| FPM | Food production model . |
| AIFI | Evolutionary algorithm for the indicator calculation of the food industry |
| UEM | An upwelling ecosystem model . |
| MWEL | Model of the typical water ecosystem on the land . |
| AHIS | An algorithm for the human indicator survivability calculation |
| DMEP | Dynamic model of the environmental pollutants. |
| BDP | The big data processing with the use of sequential and cluster analyses |
| GSA | The GIMS structure adaptation to the simulation experiment conditions |
| DFM | Big data cloud formation and management. |
| SS | Synthesis of the scenarios for the interaction of population with the environment. |
| SEMC | Simulation experiment management and control. |

## 4. Examples of global big data processing

Interaction level of society and nature has reached global planetary scales when anthropogenic impacts on the natural subsystems and processes become the dangerous changes of the habitat both for animals and peoples. There is only a single approach to the search for CNSS sustainable development, which consists in the evaluation of the consequences of anthropogenic scenario realization by means of simulation experiments. The GIMS approach can help in these experiments allowing the estimation of environmental consequences from anthropogenic scenarios realization, including global and regional scales of the impacts. Variety of these impacts covers practically all environments including pollution of the atmosphere and hydrosphere at the expense of the release of toxic compounds into the environment, the destruction of habitats through agriculture and urban sprawl, agricultural expansion to the forested areas, etc. In other words, human society aiming at living comfort, depletes the resources, destroys the vegetable kingdom and animal word, and pollutes the environment. To demonstrate the GIMS functions, several scenarios of the forest areas reconstruction are considered. Mainly the GIMS could realize different scenarios that are harmonized with our understanding of possible effects on the environment.

The land cover is characterized by the heterogeneity of biomes and other environmental objects. At present the main areas of the land are woodland (41.2%) and cropland (24.7%). The land-use is a basic problem and the solution of which involves the management and modification of the natural environment which mainly relates to the reduction of the forest areas[56]. Unfortunately, the forest areas are subject to the realistic implementation of anthropogenic scenarios associated with the withdrawal of their biomass.

It is known that boreal and tropical forest undergo the most anthropogenic impacts. The phenomenon of wildfire by lightning strike or by human actions is the primary determinant of the forests. The GIMS allows for the realization of different scenarios of influence on these forests with spatial resolution 1°×1° [57].

Simulation experiments show that total burning of all coniferous forests up to 42°N leads to an increase of atmospheric carbon by 21.7% with increasing global temperature by 4°C. During the next year, the oceans absorb 10% of the emitted carbon and during the next century only 1.3% of the increased atmospheric carbon has remained in the atmosphere. The system atmosphere-ocean over 100 years goes to the carbon balance and carbon content in the deep ocean increase by 9%. Realized impact on the coniferous forests is reflected in the humus layer dynamics, that over 30 years, has lost 4% of its reserves. Fired coniferous forests are restored during 100 years by 68%. Other forests absorb 1/30 part of emitted carbon.

Tables 3-5 show the estimates of variations of carbon reserves in basic biospheric reservoirs when forests of different climatic zones are partially burned. It is seen that large-scale impacts on the land biota are damped during 60-100 years. Under this biosphere is more stable to the impacts on the tropical forests than on boreal forests. Simulation results show that the forests of the northern hemisphere (42°N and higher) play a significant stabilization role in the global carbon cycle. Within these scenarios it is assumed that the forest areas during post-fire restoration covered by the same plants. Certainly, post-fire restoration leads to a new forest structure and other species diversity which is the result of many environmental processes including global climate change and natural forest succession.

**Table 3**. The GIMS evaluations of the variations in the carbon reserve under conditions when the coniferous forests of the Northern latitudes (up to 42°N) are burned. Nomination for $CO_2$ concentration: $C_A$ –atmosphere, $C_U$ – upper photic ocean layer, $C_S$ – dead organic matter on the land, $C_I$ – intermediate photic ocean layer under the thermocline, $C_D$ – deep ocean, and $C_B$ – near bottom ocean layer, $C_X=C_I+C_D+C_B$.

| Years after influence | Deviation in the carbon reserve (Gt) | | | | Years after influence | Deviation in the carbon reserve (Gt) | | | |
|---|---|---|---|---|---|---|---|---|---|
| | $\Delta C_A$ | $\Delta C_S$ | $\Delta C_U$ | $\Delta C_X$ | | $\Delta C_A$ | $\Delta C_S$ | $\Delta C_U$ | $\Delta C_X$ |
| Scenario: Coniferous forests are burned partly on the 25% of the area. | | | | | | | | | |
| 0 | 37.71 | -1.44 | 3.95 | 0.02 | 60 | 6.31 | -7.53 | 2.15 | 3.24 |
| 10 | 28.54 | -8.53 | 7.71 | 0.78 | 70 | 4.23 | -6.41 | 1.63 | 3.32 |
| 20 | 20.82 | -10.91 | 5.43 | 1.85 | 80 | 2.82 | -5.39 | 1.32 | 3.45 |
| 30 | 15.87 | -10.86 | 4.82 | 2.12 | 90 | 1.41 | -4.44 | 1.16 | 3.47 |
| 40 | 11.63 | -10.17 | 3.45 | 2.61 | 100 | 0.52 | -3.67 | 0.72 | 3.41 |
| 50 | 9.21 | -8.42 | 3.01 | 2.93 | 200 | -2.22 | -0.83 | -0.25 | 3.38 |
| Scenario: Coniferous forests are burned partially on the 50% of the area. | | | | | | | | | |
| 0 | 75.12 | -2.56 | 6.86 | 0.05 | 60 | 12.85 | -14.89 | 4.33 | 6.81 |
| 10 | 56.15 | -16.71 | 16.32 | 1.54 | 70 | 8.28 | -12.78 | 3.07 | 6.29 |
| 20 | 39.52 | -22.54 | 10.53 | 3.61 | 80 | 5.24 | -8.89 | 2.44 | 6.79 |
| 30 | 29.62 | -18.87 | 9.7 | 4.42 | 90 | 2.83 | -8.56 | 1.93 | 7.04 |
| 40 | 24.13 | -19.53 | 7.19 | 5.08 | 100 | 1.09 | -7.29 | 1.31 | 6.64 |
| 50 | 16.54 | -14.53 | 2.99 | 5.6 | 200 | -4.35 | -1.65 | -0.45 | 6.62 |
| Scenario: Coniferous forests are burned totally. | | | | | | | | | |
| 0 | 150.05 | -5.72 | 15.79 | 0.11 | 60 | 24.79 | -30.47 | 7.75 | 13.15 |
| 10 | 114.63 | -34.14 | 30.84 | 3.13 | 70 | 16.67 | -25.71 | 5.89 | 13.53 |
| 20 | 83.91 | -44.27 | 22.56 | 7.42 | 80 | 11.62 | -21.53 | 5.17 | 14.08 |
| 30 | 64.44 | -43.91 | 18.84 | 8.55 | 90 | 5.89 | -17.91 | 4.33 | 14.33 |
| 40 | 47.82 | -40.72 | 13.96 | 10.42 | 100 | 2.23 | -14.87 | 2.81 | 13.98 |
| 50 | 35.12 | -33.88 | 12.14 | 11.74 | 200 | -8.45 | -3.49 | -0.94 | 13.56 |

**Table 4.** Model evaluations of the variations in the carbon reserve when all the forests of the northern latitudes up to 42°N are partially burned. Nominations are given in Table 3.

| Years after impact | Deviation in the carbon reserve (Gt) | | | | Years after impact | Deviation in the carbon reserve (Gt) | | | |
|---|---|---|---|---|---|---|---|---|---|
| | $\Delta C_A$ | $\Delta C_S$ | $\Delta C_U$ | $\Delta C_X$ | | $\Delta C_A$ | $\Delta C_S$ | $\Delta C_U$ | $\Delta C_X$ |
| Scenario: 25% of the northern (up to 42°N) forests are burned. | | | | | | | | | |
| 0 | 59.52 | -1.9 | 6.22 | 0.03 | 60 | 11.05 | -12.13 | 3.54 | 5.34 |
| 10 | 43.02 | -7.64 | 11.97 | 1.23 | 70 | 8.14 | -10.31 | 2.52 | 5.41 |
| 20 | 33.75 | -16.42 | 9.8 | 2.49 | 80 | 5.23 | -8.71 | 1.91 | 5.88 |
| 30 | 27.08 | -22.03 | 7.95 | 3.45 | 90 | 3.11 | -7.49 | 1.43 | 5.89 |
| 40 | 19.68 | -15.75 | 5.78 | 4.05 | 100 | 1.81 | -6.17 | 0.91 | 6.02 |
| 50 | 15.04 | -13.99 | 4.47 | 4.83 | 200 | -3.17 | -1.48 | -9.43 | 5.42 |
| Scenario: 50% of the northern (up to 42°N) forests are burned. | | | | | | | | | |
| 0 | 119.05 | -3.95 | 13.47 | 0.54 | 60 | 23.95 | -25.67 | 6.53 | 10.82 |
| 10 | 86.92 | -15.35 | 22.56 | 2.74 | 70 | 16.48 | -20.59 | 5.35 | 11.17 |
| 20 | 72.36 | -33.59 | 18.14 | 4.98 | 80 | 10.82 | -17.85 | 3.68 | 12.31 |
| 30 | 51.68 | -41.17 | 16.48 | 6.61 | 90 | 6.17 | -17.45 | 2.97 | 12.48 |
| 40 | 41.98 | -32.92 | 11.06 | 8.49 | 100 | 3.94 | -12.32 | 2.07 | 12.87 |
| 50 | 30.32 | -29.13 | 8.81 | 9.99 | 200 | -6.6 | -3.06 | -0.79 | 10.81 |
| Scenario: 100% of the northern (up to 42°N) forests are burned. | | | | | | | | | |
| 0 | 239.02 | -7.94 | 25.18 | 0.08 | 60 | 43.78 | -48.91 | 15.67 | 20.71 |
| 10 | 174.28 | -31.62 | 48.07 | 4.93 | 70 | 33.14 | -42.18 | 10.43 | 21.93 |
| 20 | 138.91 | -67.64 | 39.27 | 11.01 | 80 | 20.72 | -35.03 | 7.54 | 22.82 |
| 30 | 107.39 | -90.33 | 32.24 | 13.82 | 90 | 12.91 | -30.12 | 5.52 | 23.33 |
| 40 | 83.08 | -64.32 | 24.11 | 16.89 | 100 | 7.33 | -24.74 | 3.71 | 23.64 |
| 50 | 61.55 | -56.92 | 17.89 | 19.21 | 200 | -12.74 | -5.94 | -1.74 | 21.73 |

**Table 5.** Model estimation of the variations in the carbon reserve when the forests of the tropical latitudes are partially burned. Nominations are given in Table 3.

| Years after influence | Deviation in the carbon reserve (Gt) | | | | Years after influence | Deviation in the carbon reserve (Gt) | | | |
|---|---|---|---|---|---|---|---|---|---|
| | $\Delta C_A$ | $\Delta C_S$ | $\Delta C_U$ | $\Delta C_X$ | | $\Delta C_A$ | $\Delta C_S$ | $\Delta C_U$ | $\Delta C_X$ |
| Scenario: 25% of tropical forests are burned. | | | | | | | | | |
| 0 | 406,2 | -20,0 | 42,2 | 0,2 | 60 | 2,9 | -12,4 | 3,0 | 22,8 |
| 10 | 264,4 | -93,7 | 74,1 | 8,0 | 70 | -5,8 | -7,5 | 0,5 | 22,8 |
| 20 | 162,2 | -84,8 | 48,0 | 14,9 | 80 | -11,6 | -4,2 | -0,9 | 22,6 |
| 30 | 90,6 | -38,6 | 27,9 | 19,1 | 90 | -13,2 | -2,6 | -1,7 | 22,5 |
| 40 | 45,4 | -36,5 | 15,0 | 21,3 | 100 | -14,5 | -1,9 | -2,1 | 21,8 |
| 50 | 18,3 | -21,6 | 7,5 | 22,4 | 200 | -13,2 | -2,3 | -1,9 | 17,7 |
| Scenario: 50% of tropical forests are burned. | | | | | | | | | |
| 0 | 406,2 | -20,0 | 42,2 | 0,2 | 60 | 2,9 | -12,4 | 3,0 | 22,8 |
| 10 | 264,4 | -93,7 | 74,1 | 8,0 | 70 | -5,8 | -7,5 | 0,5 | 22,8 |
| 20 | 162,2 | -84,8 | 48,0 | 14,9 | 80 | -11,6 | -4,2 | -0,9 | 22,6 |
| 30 | 90,6 | -38,6 | 27,9 | 19,1 | 90 | -13,2 | -2,6 | -1,7 | 22,5 |
| 40 | 45,4 | -36,5 | 15,0 | 21,3 | 100 | -14,5 | -1,9 | -2,1 | 21,8 |
| 50 | 18,3 | -21,6 | 7,5 | 22,4 | 200 | -13,2 | -2,3 | -1,9 | 17,7 |
| Scenario: 100% of tropical forests are burned. | | | | | | | | | |
| 0 | 406.2 | -20.0 | 42.2 | 0.2 | 60 | 2.9 | -12.4 | 3.0 | 22.8 |
| 10 | 264.4 | -93.7 | 74.1 | 8.0 | 70 | -5.8 | -7.5 | 0.5 | 22.8 |
| 20 | 162.2 | -84.8 | 48.0 | 14.9 | 80 | -11.6 | -4.2 | -0.9 | 22.6 |
| 30 | 90.6 | -38.6 | 27.9 | 19.1 | 90 | -13.2 | -2.6 | -1.7 | 22.5 |
| 40 | 45.4 | -36.5 | 15.0 | 21.3 | 100 | -14.5 | -1.9 | -2.1 | 21.8 |
| 50 | 18.3 | -21.6 | 7.5 | 22.4 | 200 | -13.2 | -2.3 | -1.9 | 17.7 |

Table 6 shows the change of vegetation role in the atmospheric carbon absorption under the reconstruction of soil-plant formations. Anthropogenic change of vegetation covers can significantly change balance components of global carbon cycle. It is clear that such hypothetical vegetation cover transformations need to take into account climatic zones and biological compatibility. The GIMS partly helps to realize similar simulation experiments.

**Table 6.** The dynamics of the ratio of the integral rates of $CO_2$ assimilation by vegetation covers from the atmosphere in the context of the scenario when natural biome is globally replaced by other biome.

| Scenario when the actual biome is replaced by another biome globally in 1-degree scale. Scenario is realized in 2015. | | Coefficient of the rate of $CO_2$ assimilation change. | | | |
|---|---|---|---|---|---|
| | | Years | | | |
| | | 2020 | 2030 | 2050 | 2100 |
| Really existing biome | Replacing biome | | | | |
| Arctic deserts and tundra | Forest-tundra | 2.5 | 2.1 | 2.0 | 2.2 |
| Tundras | Forest-tundra | 0.9 | 0.9 | 1.0 | 1.0 |
| Mountain tundra | Forest tundra | 1.4 | 1.2 | 1.0 | 1.0 |
| North-taiga forests | Mid-taiga forests | 1.6 | 1.4 | 1.1 | 1.1 |
| South-taiga forests | Mid-taiga forests | 1.9 | 1.7 | 1.5 | 1.3 |
| Broad-leaved coniferous forests | Mid-taiga forests | 3.9 | 3.7 | 2.9 | 2.0 |
| Sub-tropical broad-leaved and coniferous forests | Humid evergreen tropical forests | 3.1 | 2.6 | 2.4 | 1.8 |
| Xerophytic open woodlands and shrubs | Humid evergreen tropical forests | 21.5 | 19.5 | 17.9 | 18.3 |
| Moderate arid and arid (mountain including) steppes | Humid evergreen tropical forests | 22.2 | 18.3 | 16.3 | 14.8 |
| Pampas and grass savannas | Humid evergreen tropical forests | 99.2 | 77.1 | 68.5 | 70.1 |
| Sub-tropical deserts | Humid evergreen tropical forests | 187.7 | 153.2 | 138.4 | 140.7 |
| Alpine and sub-alpine meadows | Humid evergreen tropical forests | 790.0 | 766.4 | 751.3 | 767.3 |
| Variabli-humid deciduous tropical forests | Humid evergreen tropical forests | 1.4 | 1.3 | 1.2 | 1.3 |
| Tropicalxerophytic open woodlands | Humid evergreen tropical forests | 67.6 | 60.2 | 56.6 | 57.3 |
| Tropical savannas | Humid evergreen tropical forests | 5.9 | 5.0 | 4.7 | 5.1 |
| Tropical deserts | Humid evergreen tropical forests | 25.9 | 24,8 | 23.6 | 22.6 |
| Sub-tropical and tropical grass-tree thickets of the tugai type | Humid evergreen tropical forests | 17.1 | 15.8 | 14.9 | 14.0 |
| Sub-tropical semi-deserts | Humid evergreen tropical forests | 0.9 | 1.2 | 0.9 | 1.1 |

The GIMS gives the opportunity to evaluate the mosaic picture of the carbon dioxide sinks in the vegetation biomes in its dynamics. The knowledge of this mosaic makes it possible to assess the role of concrete biomes in the regional balance of carbon and on this basis, to estimate the possible consequences from anthropogenic interference in these biomes. The GIMS gives an opportunity to estimate the atmospheric $CO_2$ sequestration by vegetation sites in their evolution on different territories. It is assumed that $CO_2$ emissions in 2015 are estimated by 36.1 $GtCO_2$ globally and 1.7 $GtCO_2$ from the Russian territory with consecutive decrease by 10 percents to 2150. As well as it is accepted that deforestation processes on Russian territory are no realized. In the context of this scenario, rates of $CO_2$ assimilation by plants on the territory of Russia will increase from 206.1 MtC/year in 2015 to 292.3 MtC/year in 2150 which is a result of climate change. Under this role of arctic deserts and tundras are characterized by the increase from 2000 to 2150 by 256%. Global warming effects on these biomes are shown in the expansion of the vegetation period and transformation plant types at the expense of the tundra-taiga boundary succession processes. Such effects are reduced for the mid-taiga forests and dry steppes up to 205% and 114%, respectively.

Natural land cover transformations are real processes that are realized by humans for the improvement of living habitat and food production growth. These actions lead to the change of many evolutionary processes, including changes in biogeochemical cycles, which directly leads to climate change. Table 6 demonstrates some modeling results when various biomes are transformed. Such hypothetical experiments lead to the understanding of the limits of natural stability and to the possible ranges of anthropogenic interventions.

Simulation experiments show significant dependence of global climate on the overall health of global forests. For example, the reduction of global forested areas by 19% to 2050 leads to the $CO_2$ concentration increase by 53% by the end of 21$^{st}$ century and, on the contrary, increase of forested areas by 10% during the same time period gives a $CO_2$ concentration decrease by 12%. Figure 5 represents a climate change within realization of different scenarios for the impacts on forests. We see that forests and climate are intrinsically linked significantly through the sequestering atmospheric carbon as well as through direct and indirect impacts on the global hydrological cycle[58].

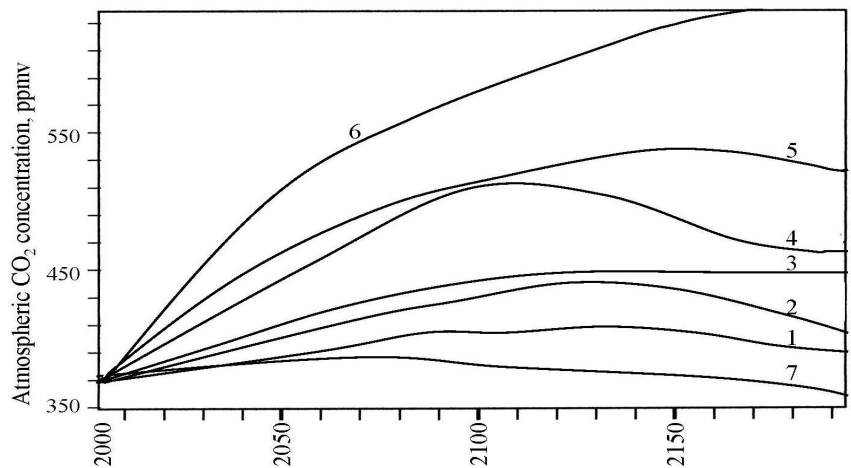

**Figure 5.** The dynamics of the $CO_2$ concentration for different scenarios of changing forest areas in the context of possible scenarios: 1 by 2050 the area of the forests is increased by 5% and remains without changes; 2 by 2050 the area of the forests is reduced by 5%; 3 by 10%; 4 by 20%; 5 by 30%; 6 by 2050 the forests will be liquidated at all; 7 by 2050 the area of the forests is increased by 50% without change in the future.

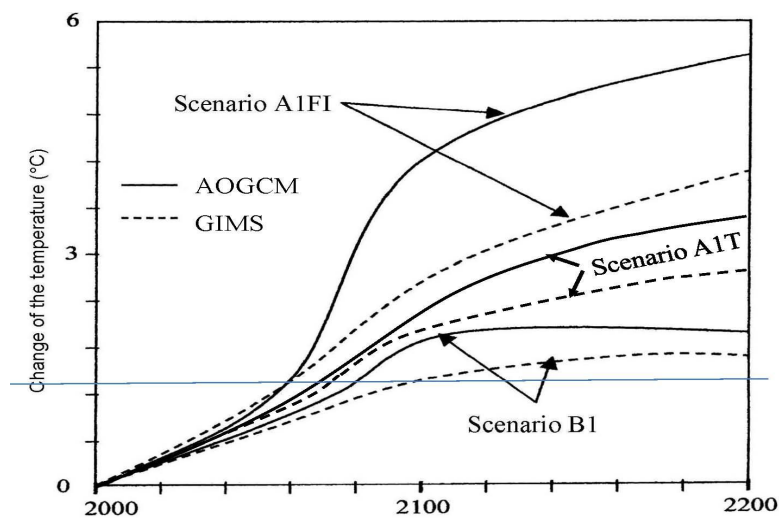

**Figure 6.** Prognoses of the changes in average global temperature evaluated by means of AOGCM and GIMS within three IPCC emission scenarios.

The GIMS reflects the interactions of natural and anthropogenic factors that play a significant role in the greenhouse effect formation depending on the energy use and economic development. Figure 6 gives a comparison of modeling results regarding the future global temperature change under the realization of IPCC scenarios. It is evident that the GIMS forecasts more low deviations in average global temperature compared to results from the atmosphere-ocean general circulation model (AOGCM) of Hadley Centre. For example, the realization of the A1FI pessimistic scenario is geared to a very rapid economic growth and intensive use of fossil resources, giving the global temperature rise to 2100 by 4°C and 2.6°C according to AOGCM and GIMS, respectively. But these changes to 2200 are 5.5°C and 4°C, respectively. The conclusion that can be drawn from the results of Fig. 6 is that GIMS gives more precise results due to more broad components taking into consideration.

## 5. Conclusions

Co-evolution of climate, biosphere, geosphere, hydrosphere and human society depends on how the Earth's system generates and maintains thermodynamic imbalance. Understanding and evaluating processes in the climate-nature-society system requires the big data processing algorithms under the exponential growth of them and when using traditional data processing tools eventually become obsolete. Most of the existing climate models and global biospheric models do not provide overall analysis of the processes existent in the Earth system. The GIMS as it is seen in Table 2 and Figure 1 can play the role of the Big Data information-modeling system that at one time can analyze heterogeneous data delivered by different monitoring systems with incongruous scales and un-removable uncertainties. Tables 3-6 demonstrate such functions of the GIMS as a new Big Data approach.

As a result of simulation results, there are significant impacts of the saturation of greenhouse effect due to $CO_2$ growth, which is agreed with the physical laws and is justified by many modeling results. Simulation experiments analogous to Table 6 can help to search for optimal forest management strategy when climate change forecast will be acceptable for a long time. Certainly, this paper can not solve this task. It is necessary to realize series of simulation experiments with consideration of reasonable scenarios.

Moreover, the GIMS possesses the data fusion function when data are delivered from dissimilar sources by irregularly in time and fragmentary by space. This function allows for the answer on the following questions that inevitable arisen under environmental monitoring management:
- What tools, remote-sensing platforms and instruments should be used to form the GIMS database?
- What is the cost of information for the GIMS simulation experiment?
- What balance should be between different data sources?

In this paper we have been able to show the way of addressing these and other questions, that are predominately linked to CNSS sustainable development using simple biosphere models and complex simulation models that require the big data processing. In any case, the search for alternative pathways towards global CNSS sustainable development is realized through simulation experiments in the context of scenarios that are proposed by experts. The GIMS extends the sphere of scenarios and optimizes the big data fluxes.

This paper describes the main structure of GIMS and gives examples of its use to demonstrate functional efficiency of the big data analysis and processing. Scenarios that are studied here show the presence of alternatives in environmental anthropogenic strategies.